\documentclass[lettersize,journal]{IEEEtran}
\usepackage{amsmath,amsfonts}
\usepackage{algorithmic}
\usepackage{algorithm}
\usepackage{array}
\usepackage[caption=false,font=normalsize,labelfont=sf,textfont=sf]{subfig}
\usepackage{textcomp}
\usepackage{stfloats}
\usepackage{url}
\usepackage{verbatim}
\usepackage{graphicx}
\usepackage{cite}
\usepackage[colorlinks, citecolor=green]{hyperref}

\begin{document}

\title{ROI-based Deep Image Compression with Implicit Bit Allocation}

\author{Kai Hu, 
        Han Wang,
        Renhe Liu,
        Zhilin Li,
        Shenghui Song,
        Yu Liu*
\thanks{Kai Hu, Han Wang, Renhe Liu, Zhilin Li, and Yu Liu are with the School of Microelectronics, Tianjin University, Tianjin 300072, China (email: kaihu@tju.edu.cn).}
\thanks{Shenghui Song is with the Hong Kong University of Science and Technology, Hongkong, China.}
\thanks{Yu Liu is the corresponding author (email: liuyu@tju.edu.cn).}

\markboth{Journal of \LaTeX\ Class Files,~Vol.~14, No.~8, August~2025}%
{Shell \MakeLowercase{\textit{et al.}}: A Sample Article Using IEEEtran.cls for IEEE Journals}
}


\maketitle

\begin{abstract}
Region of Interest (ROI)-based image compression has rapidly developed due to its ability to maintain high fidelity in important regions while reducing data redundancy. However, existing compression methods primarily apply masks to suppress background information before quantization. This explicit bit allocation strategy, which uses hard gating, significantly impacts the statistical distribution of the entropy model, thereby limiting the coding performance of the compression model. In response, this work proposes an efficient ROI-based deep image compression model with implicit bit allocation.
To better utilize ROI masks for implicit bit allocation, this paper proposes a novel Mask-Guided Feature Enhancement (MGFE) module, comprising a Region-Adaptive Attention (RAA) block and a Frequency-Spatial Collaborative Attention (FSCA) block. This module allows for flexible bit allocation across different regions while enhancing global and local features through frequency-spatial domain collaboration. Additionally, we use dual decoders to separately reconstruct foreground and background images, enabling the coding network to optimally balance foreground enhancement and background quality preservation in a data-driven manner. 
To the best of our knowledge, this is the first work to utilize implicit bit allocation for high-quality region-adaptive coding. 
Experiments on the COCO2017 dataset show that our implicit-based image compression method significantly outperforms explicit bit allocation approaches in rate-distortion performance, achieving optimal results while maintaining satisfactory visual quality in the reconstructed background regions.

\end{abstract}


\section{Introduction}
\IEEEPARstart{R}{egion} of Interest (ROI)-based image compression leverages a fundamental characteristic of human visual perception: viewers’ attention is not uniformly distributed across an image. Instead, it tends to concentrate on one or several salient regions, often associated with prominent foreground objects. Consequently, in ROI-based coding schemes, these foreground objects are automatically segmented as ROIs and allocated more bits compared to background regions. This strategy not only preserves critical visual details in key areas but also facilitates higher compression ratios. By reducing data redundancy in image processing and visual analysis tasks, ROI-based image compression enhances storage efficiency and ensures reliable transmission of semantically important content~\cite{tahoces2008image}. It is particularly valuable in bandwidth-constrained environments such as telemedicine, video surveillance, and mobile multimedia communications.

\begin{figure}[!t] 
	\centering    
	\includegraphics[scale=0.75]{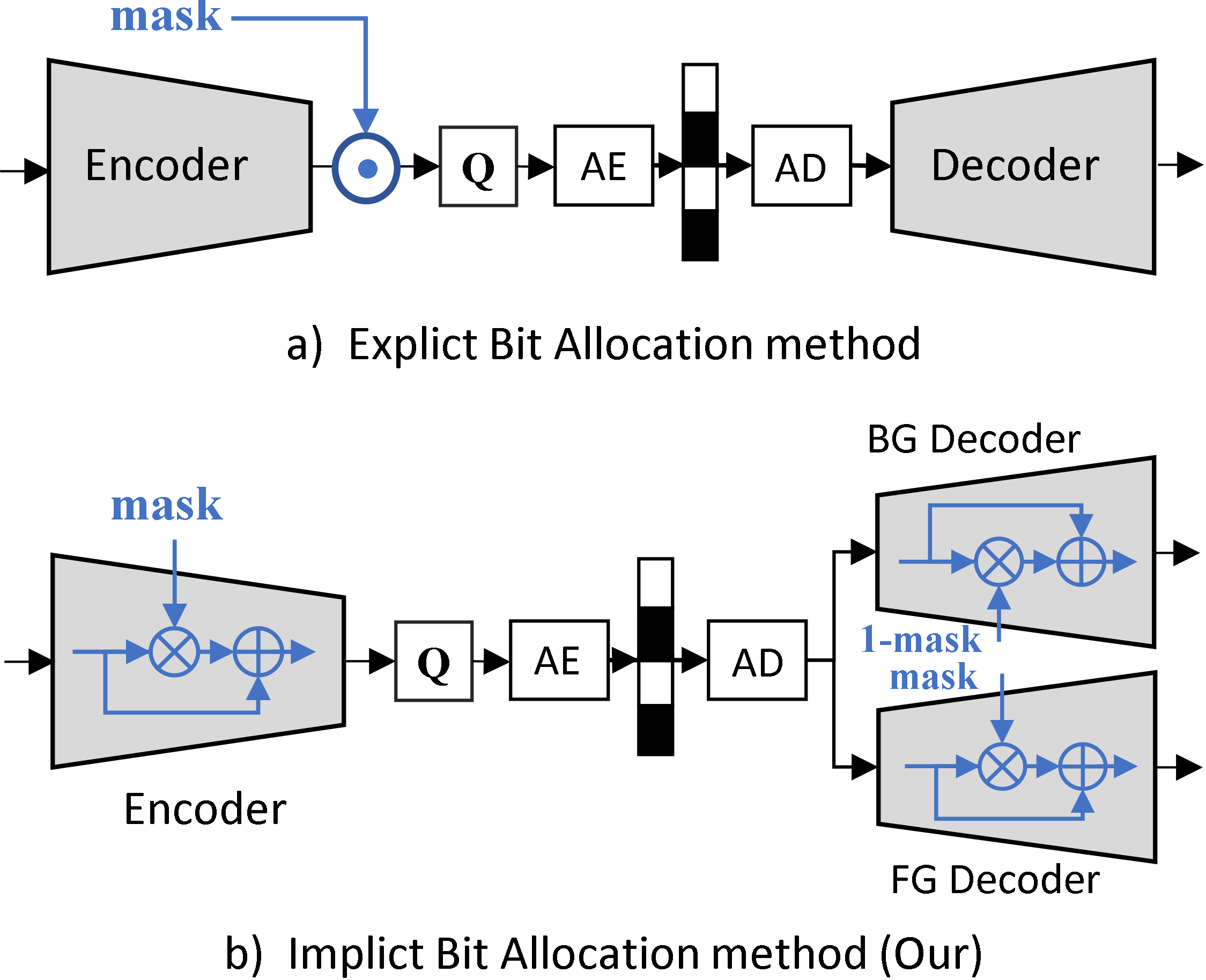} 
	\caption{An overview of different bit allocation methods. Our approach not only utilizes ROI masks to generate attention maps for adaptively enhancing or suppressing features across different regions but also employs a dual-decoder architecture to balance the reconstruction quality between the foreground (FG) and background (BG).}
	\label{fig:bitAllo_comp}
\end{figure}
With the advancement of convolutional neural networks (CNNs), ROI-based image compression has rapidly progressed due to its superior coding performance. As noted by Li et al.~\cite{li2024saliency}, existing ROI-based deep compression methods can be broadly categorized into two groups based on their bit allocation strategies: explicit and implicit approaches. For explicit bit allocation, Li et al.~\cite{li2018learning, li2020learning} proposed a mapping subnet to generate importance maps for key regions, enabling adaptive bit allocation based on image content. Similarly, Mentzer et al.~\cite{mentzer2018conditional} employed importance maps to guide bit allocation. Based on~\cite{cai2018efficient}, Cai et al.~\cite{cai2019end} utilized ROI masks to perform dot products with multiscale feature representations. Ma et al.~\cite{ma2021variable} applied 3D masks to prune latent representations, improving background bit allocation. Their method also incorporated Generative Adversarial Networks (GANs) to supervise the training of the entire coding network, thereby enhancing the quality of foreground reconstruction. Li et al.~\cite{li2024saliency} decomposed latent features into base and enhancement layers, then applied a mask to zero out background regions in the enhancement layer. They used a dual-scale entropy module for separate entropy coding of different feature layers.

In these works, explicit bit allocation primarily involves masking the feature representation before quantization and suppressing background information through hard gating to achieve bit allocation, as shown in Fig.~\ref{fig:bitAllo_comp}(a). However, this strategy disrupts the statistical distribution of the entropy model and limits the compression performance of the network. For implicit bit allocation, Cheng et al.~\cite{cheng2020learned},  Zou et al.~\cite{zou2022devil} and Lu et al.~\cite{lu2022high} introduced attention modules to capture global correlations within images. By computing these global correlations to generate attention maps for implicit bit allocation, more complex regions—which are harder to reconstruct—are assigned greater weights in the attention maps, thereby preserving more detailed information. Nevertheless, these implicit methods still cannot achieve ROI-based image compression.

To address the above issues, this paper proposes an efficient ROI-based deep image compression model with implicit bit allocation, as shown in Fig.~\ref{fig:bitAllo_comp}(b). To achieve adaptive bit allocation, we introduce a simple region-adaptive attention (RAA) block that utilizes masks to generate attention maps. The entire RAA block is smooth and fully differentiable, facilitating end-to-end joint training and optimization of the coding network. We further propose a Frequency-Spatial Collaborative Attention (FSCA) block to enhance the representation of weighted feature maps globally and locally. By integrating the RAA and FSCA blocks, we develop a novel Mask-Guided Feature Enhancement (MGFE) module, which guides the coding network's learning process progressively across layers. Since our implicit method retains the complete spatial context, we employ dual decoders to reconstruct the foreground and background separately. This allows the coding network to learn an optimal balance between enhancing ROI and preserving background quality in a data-driven manner.

Our contributions are summarized as follows:
\begin{itemize}
\item This paper proposes an efficient ROI-based deep image compression model with implicit bit allocation. To the best of our knowledge, this is the first work to employ implicit bit allocation to achieve high-quality region-adaptive coding.
\item To apply ROI masks for implicit bit allocation, we introduce a novel MGFE module, which consists of an RAA block and a FSCA block. This module enables flexible bit allocation across different regions and enhances features both globally and locally through a collaboration between the frequency and spatial domains.
\item We provide a detailed analysis of implicit versus explicit bit-allocation strategies. Our experimental results demonstrate that our implicit-based approach achieves superior coding performance compared to explicit methods by incorporating masks in a smooth and differentiable manner.
\item Our method achieves excellent rate-distortion (RD) performance while maintaining an optimal balance between improving ROI and preserving background quality. Notably, our compressed images also yield the highest recognition accuracy in computer vision tasks, even outperforming the original uncompressed images.
\end{itemize}

\section{RELATED WORKS}
\subsection{Learned Image Compression}
Ballé et al.~\cite{balle2018variational} first developed an end-to-end learned image compression (LIC) model by replacing the quantization with additive uniform noise and utilizing a learnable GDN/IGDN. To enhance the codec's ability to model image features, many researchers~\cite{9765370,cheng2020learned,10564141,9385968} have introduced residual blocks and non-local attention mechanisms. Xie et al.~\cite{xie2021enhanced} replaced the VAE network with invertible neural networks (INNs) to improve the model's coding performance. They also employed an attentive channel squeeze layer to address the instability and sub-optimal training of INN-based networks. Thanks to the Swin-Transformer's strength in capturing global feature correlations, Lu et al.~\cite{lu2022transformer} and Zhu et al.~\cite{zhu2021transformer} introduced the Swin-Transformer~\cite{liu2021swin} block into the LIC task to enhance the non-linear representation capability of the transform network. Zou et al.~\cite{zou2022devil} explored the relationship between local attention mechanisms and the global structure learned by neural networks, proposing a window-based attention block to capture the correlation between spatially adjacent elements. Lu et al.~\cite{lu2022high} further proposed local window-based residual neighborhood attention blocks to dynamically weight and aggregate adjacent elements, thereby improving the representation of instantaneous content. Liu et al.~\cite{liu2023learned} adopted parallel Transformer-CNN Mixture (TCM) blocks in the transform network and a Swin-Transformer-based attention module in the channel-wise entropy model to improve coding performance.

In order to achieve differentiated bit allocation in the image, some work has been carried out. Wang et al.~\cite{wang2022discretized} introduced a masked branch network to assist the encoder in achieving adaptive bit-rate allocation, which means that more bits are allocated to spatially complex regions and fewer bits are allocated to spatially uniform regions. Due to the fact that existing image compression methods rarely focus on the adaptability of models to handle images with different contexts or distributions. Wang et al.~\cite{wang2022neural} constructed a neural data-dependent transform for LIC to achieve better RD performance. Liu et al.~\cite{liu2024region} further implemented image-level transform to limit the description of signals to fine-grained levels, and designed a learnable region adaptive transform to capture the mapping patterns of different regions.

\subsection{ROI-based Image Compression}
ROI-based deep image compression fundamentally remains an RD optimization (RDO) problem, with the primary goal of minimizing the weighted sum of bit-rate and ROI distortion. In this framework, ROI distortion is penalized much more than background regions. A mathematical model is formulated to describe the optimization difference between ROI-based and ordinary image compression~\cite{cai2019end}, expressed as:
\begin{equation}
    \mathcal{F}^* = \arg \min_{\mathcal{F}} \left( \mathcal{R}(\mathcal{F}) + \lambda_{roi} D_{roi}(\mathcal{F}) + \lambda_{im} D_{im}(\mathcal{F}) \right).
    \label{eq:optimization}
\end{equation}
Where $\mathcal{F} = \{\mathcal{P}, \mathcal{C} \}$, where $\mathcal{P}$ is any predictive model that can be used to segment ROI, and $\mathcal{C}$ is an image codec used for ROI coding with a given mask. $\mathcal{R}$ is the bit rate. $D_{roi}$ is the ROI distortion, and $D_{im}$ is the average distortion of the entire image. Among them, $\mathcal{R}$, Both $D_{roi}$ and $D_{im}$ are related to the performance $\mathcal{F}$ of the designed codec. $\lambda_{im}$ and $\lambda_{roi}$ are Lagrangian parameters that balance between rate and distortion. There is $\lambda_{roi}$ much larger than $\lambda_{im}$.

In recent years, ROI-based deep image compression techniques have become a new focus. Li et al.~\cite{li2018learning, li2020learning} introduced a mapping subnet for key regions to generate explicit important mappings, which adaptively allocate bits based on image contents. Mentzer et al.~\cite{mentzer2018conditional} also used importance maps to guide bit allocation. Based on~\cite{cai2018efficient}, Cai et al.~\cite{cai2019end} used ROI masks to dot product multi-scale representations, which mainly allows more bits to be allocated to the foreground in the image.  Ma et al.~\cite{ma2021variable} adopted 3D masks to prune latent representations to improve bit allocation in background regions and employed GANs for supervised training of the entire coding network. Li et al.~\cite{li2023roi} integrated ROI masks into different layers of the compression network, achieving spatial adaptability and variable-rate compression by modifying the Lagrange multiplier $\lambda$ in different regions. Li et al.~\cite{li2024saliency} developed a saliency segmentation oriented deep compression model, employing a saliency segmentation model to generate masks. For bit allocation across different image regions, they decomposed latent features into base and enhancement layers prior to quantization, applying masks to locally zero-out background areas in the enhancement layer, followed by entropy coding using a Double-Scale Entropy Module. Jin et al.~\cite{jin2025customizable} introduced a customizable ROI-based deep image compression framework incorporating user-provided textual information. Their proposed framework enables both customizable ROI compression and user-adjustable quality trade-offs between ROI and non-ROI reconstruction. In these works, ROI-based image compression is achieved by applying masks to feature representations before quantization, and then partially or completely discarding background region information.

\begin{figure*}[!t] 
	\centering    
	\includegraphics[scale=0.59]{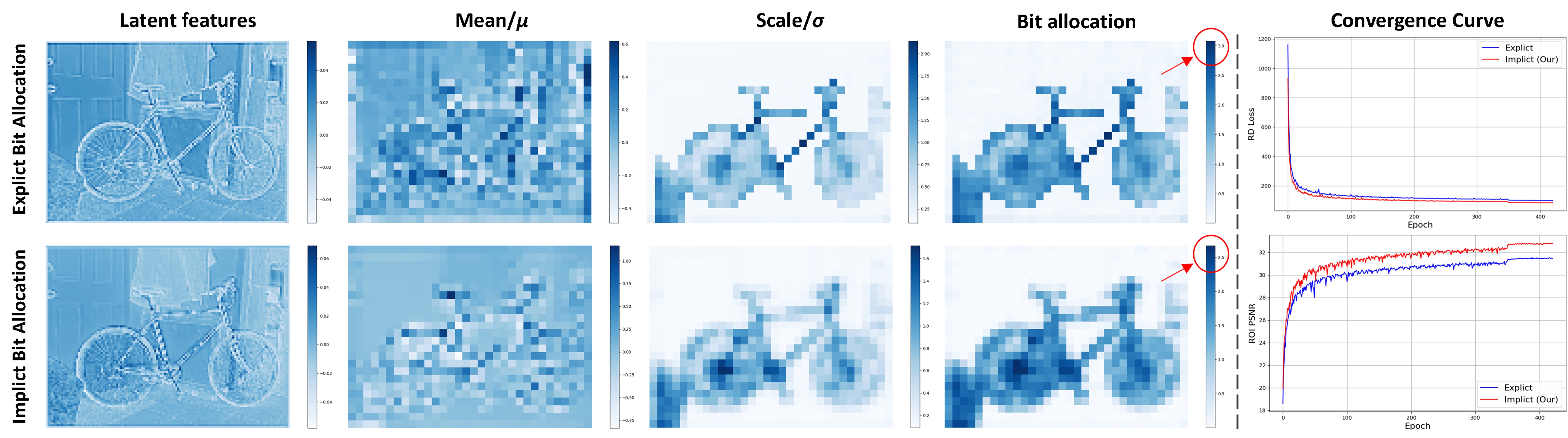} 
	\caption{Visualization of latent features and bit allocation maps trained with explicit and implicit bit allocation methods. 
    We also visualize the loss curves for RD optimization during training and the ROI-PSNR curves on the test validation set under different methods. The integrity of ROI features obtained by explicit methods is limited and the global image occupies more bits, especially ROI. In addition, it can be seen from the convergence curve that the explicit method converges too early, which can easily lead to local optima and result in suboptimal RD performance.}
	\label{fig:Im_explivisual}
\end{figure*}
\subsection{Feature learning in the Frequency Domain}
The Fourier transform is a widely used technique for analyzing frequency content in signals, which can be viewed as representing global signal statistics. Consequently, it can capture long-range dependency relationships between features within images. Building upon this principle, numerous existing works have explored frequency representation learning based on the Fast Fourier Transform (FFT). Chi et al.~\cite{Chi_2020_FFC} designed a fast Fourier convolution (FFC) operator with non-local receptive fields and cross-scale fusion within convolutional units to address the limitation of standard convolutions in coding long-range dependencies of image features. Kong et al.~\cite{kong2023efficient} developed an efficient frequency-domain self-attention solver (FSAS) by exploring transformer properties in the frequency domain, using element-wise product operations rather than spatial-domain matrix multiplication to estimate scaled dot-product attention for high-quality image deblurring. Inspired by modeling the convolution theorem through token mixing, Li et al.~\cite{li2025fouriersr} proposed a Fourier transform-based plugin (FourierSR) to reduce complexity while maintaining global receptive fields. To efficiently utilize different frequency information in images, many researchers~\cite{cui2023selective,tatsunami2024fft,gao2024exploring,9627540} investigate how to select the most informative frequency components for image reconstruction. Zhou et al.~\cite{zhou2022spatial} explored the fusion of spatial and frequency domain information by using panchromatic (PAN) images to guide the recovery of phase and amplitude information in the Fourier domain. Li et al.~\cite{li2023embedding} proposed UHDFour, a method that employs Fourier transforms through a cascaded network architecture to process ultra-high-definition images in low-light conditions, separating amplitude and phase to reduce noise. Wang et al.~\cite{wang2023spatial} constructed a dual-branch spatial and frequency mutual learning network to fully integrate facial image information from both the spatial and frequency domains. Although several existing works have explored feature fusion between frequency and spatial domains for images, their utilization of frequency-domain and spatial-domain contexts remains limited, consequently compromising the representational capacity of image information.

\begin{figure*}[!t] 
	\centering    
	\includegraphics[scale=0.64]{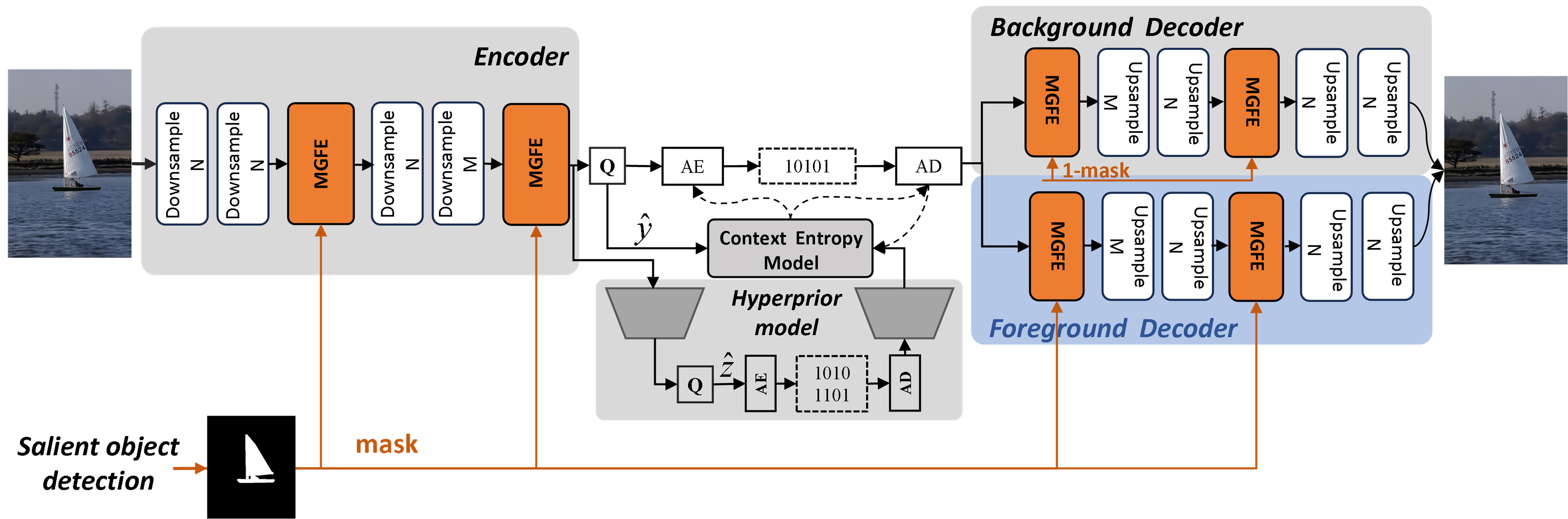} 
	\caption{The overall framework of our proposed method. ``MGFE" represents Mask-guided Feature Enhancement module. AE and AD are arithmetic en/de-coder, respectively. N is set to 192, and M is 320 as in~\cite{zou2022devil}.}
	\label{fig:network}
\end{figure*}
\section{OUR METHOD}
\subsection{Analysis of Bit Allocation Methods} 
Current ROI-based deep compression methods employ primarily an explicit bit allocation strategy~\cite{cai2019end, li2024saliency, jin2025customizable}. As shown in Fig.~\ref{fig:bitAllo_comp}(a), this strategy uses ROI masks to perform a dot product with multi-scale feature representations $\boldsymbol{y}$ before quantization, thereby achieving differential bit allocation between the foreground and background. However, this explicit approach relies on a hard gating mechanism, which leads to irreversible information erasure in the background (non-ROI) regions of the latent space. In contrast, our implicit bit-allocation method transforms a hard mask into a soft attention map to adaptively enhance the foreground and suppress the background. Crucially, this approach keeps the spatial context of the latent representation intact. Due to end-to-end learned codecs estimate bitrate by learning a probability model $p_{\hat{y}} (\boldsymbol{\hat{y}})$ that follows a Gaussian distribution (usually conditional on hyperprior and spatial context information)~\cite{cheng2020learned}. These codecs are trained on the statistical properties of latent representations derived from natural images. Therefore, the closer the distribution of the latent representation is to a Gaussian distribution, the lower the estimated bitrate.

\begin{figure}[!t] 
	\centering    
	\includegraphics[scale=0.61]{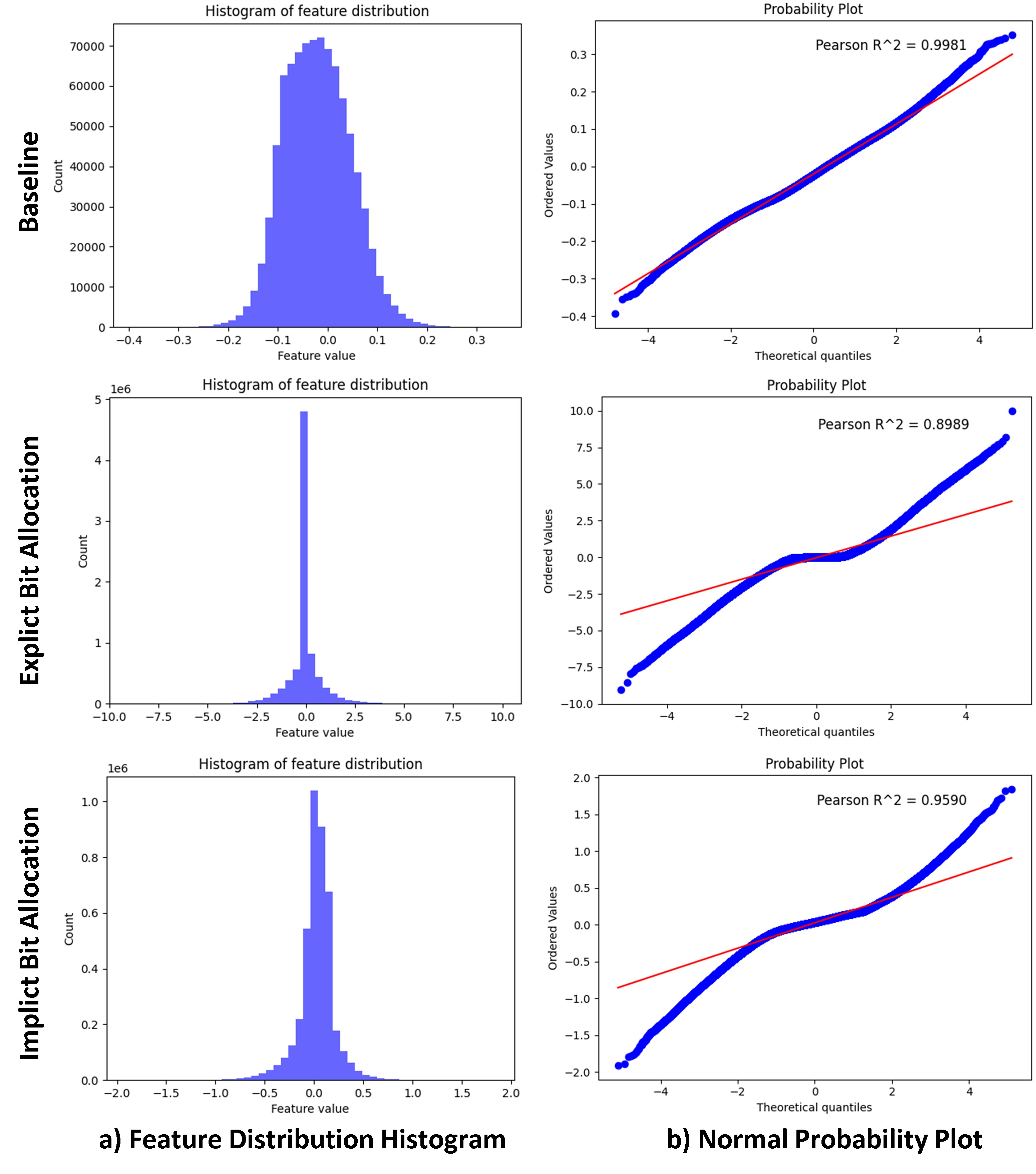} 
	\caption{The distribution histogram and normal probability plot obtained from the quantized latent features $\boldsymbol{\hat{y}}$ statistics. The experimental validation used over 3000 images from the COCO2017 dataset. The baseline model is the classic LIC model (STF)~\cite{zou2022devil}.}
	\label{fig:distrimao}
\end{figure}
For explicit bit-allocation models, we adopt the training strategy and the bit-allocation method from~\cite{cai2019end}. (Please refer to Sec.IV-D for detailed instructions.) 
To visually compare the impact of different bit allocation methods on the feature distribution and the entropy model's statistics, we visualize the feature distributions using histograms and normal probability plots, as shown in Fig.~\ref{fig:distrimao}.
A normal probability plot assesses how closely a dataset follows a Gaussian distribution; dataset from a Gaussian distribution will form an approximately straight line. For example, 
the feature distribution of the baseline model~\cite{zou2022devil} closely follows a Gaussian distribution, as shown in Fig.~\ref{fig:distrimao}b. Consequently, its histogram exhibits a standard bell-shaped curve, and the data points in its normal probability plot form an approximately straight line. We also quantify the deviation from the normal distribution by calculating the Pearson correlation coefficient (PCC) between the sample quantiles and the theoretical quantiles. A PCC value closer to 1 indicates a distribution closer to normal. Our results show that the features of our model more closely follow a Gaussian distribution than those processed by the explicit method. This is demonstrated by a PCC of 0.959 for our method, compared to 0.899 for the explicit method.

Additionally, we visualize the latent features and bit allocation maps for bit allocation methods, as shown in Fig.~\ref{fig:Im_explivisual}. Our approach achieves a more rational bit allocation while enabling smoother and more complete representation of foreground objects. In contrast, the explicit bit allocation method disrupts the entropy model's statistical distribution and severely degrades the predictive capability of spatial context models. This disruption significantly increases the overall coding cost, particularly in the foreground regions. In conclusion, developing an ROI-based compression model with implicit bit allocation is crucial to overcome the performance limitations of existing methods.

\subsection{Overview of Our Approach}
The complete compression framework proposed in this paper is illustrated in Fig.~\ref{fig:network}. We employ the network in~\cite{zou2022devil} as the backbone but replace its channel-wise auto-regressive entropy model~\cite{minnen2020channel} with a 5$\times$5 masked-convolution context model~\cite{minnen2018joint}. An original image $\boldsymbol{x} \in{\mathbb{R}}^{H \times W \times 3}$ is first fed into the compression network. Masks can be generated by pre-trained models of salient object detection~\cite{9548849}. 
To leverage the generated $\boldsymbol{mask}$ for implicit bit allocation, we introduce a well-designed MGFE module, which we integrate layer by layer in place of the initial Window Attention Module. This design precisely distinguishes foreground from background regions, enabling high-quality region-adaptive coding. Owing to our implicit bit allocation strategy, the quantized latent features $\boldsymbol{\hat{y}}$ retain intact background information. To balance the high fidelity of salient content with overall perceptual quality, we depart from previous work~\cite{cai2019end,ma2021variable,li2023roi} that uses a single decoder. Instead, we deploy dual decoders to reconstruct the foreground $\boldsymbol{\hat{x}}_f$ and background $\boldsymbol{\hat{x}}_b$ independently. The masks supplied to these two decoders are complementary, ensuring that each focuses exclusively on its designated region.
The final output $\boldsymbol{\hat{x}}$ is obtained by fusing these two region-specific outputs as 
\begin{equation}
    \begin{aligned}
        \boldsymbol{\hat{x}} = \boldsymbol{\hat{x}}_f \times \boldsymbol{mask} + \boldsymbol{\hat{x}}_b \times (\boldsymbol{1} - \boldsymbol{mask}),
    \end{aligned}
\end{equation}

\subsection{Mask-guided Feature Enhancement Module}
\begin{figure}[!t] 
	\centering    
	\includegraphics[scale=0.468]{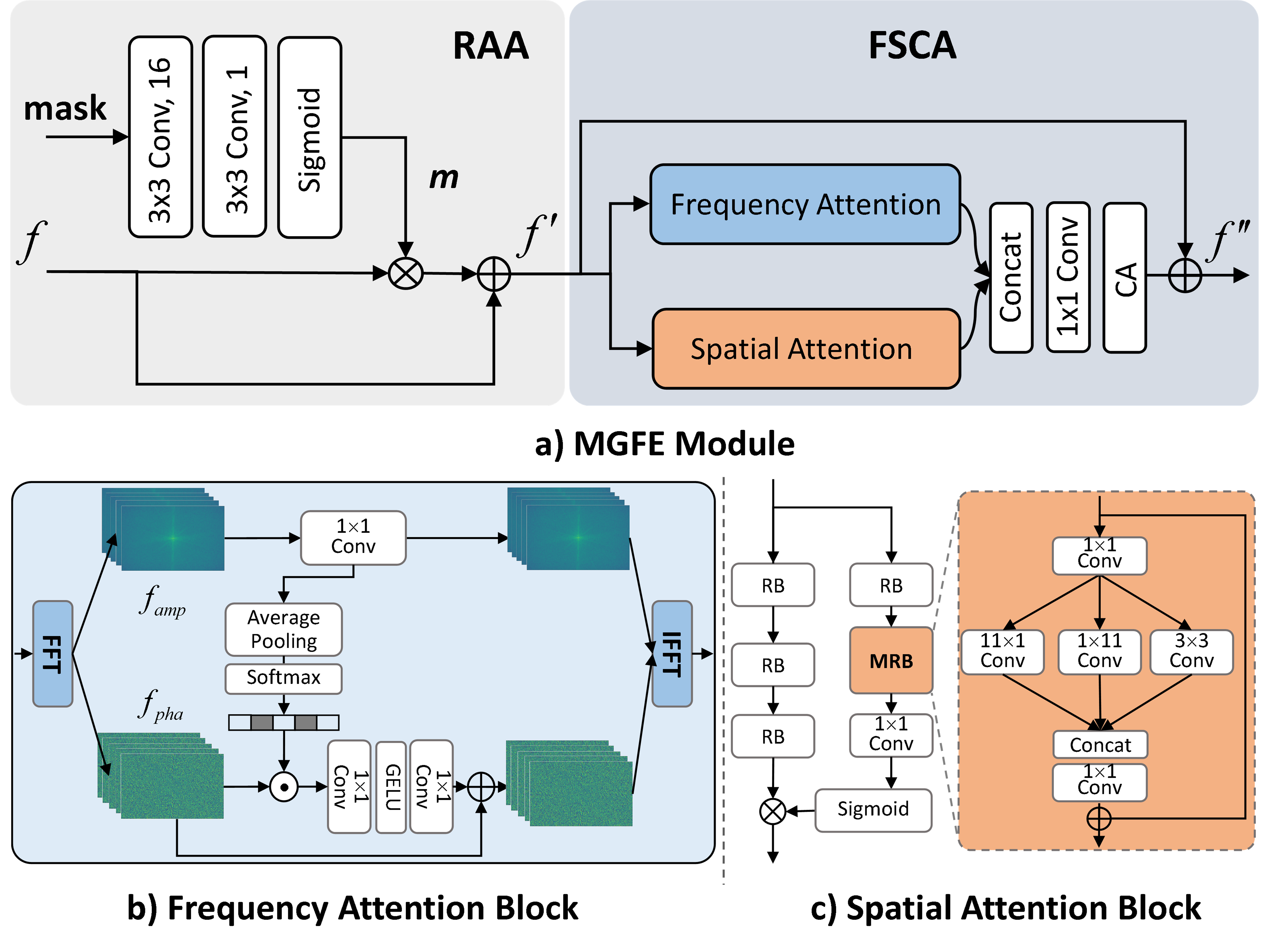} 
	\caption{Illustration of the MGFE Module, FA and SA block in FSCA block. The MGFE Module is mainly composed of RAA block and FSCA block.}
	\label{fig:MGFE}
\end{figure}

To better integrate implicit bit allocation into ROI-based compression network, we develop an efficient MGFE module, as illustrated in Fig.~\ref{fig:MGFE}a, which consists of an RAA block and an FSCA block. The RAA block utilizes the input $\boldsymbol{mask}$ to generate a dynamic spatial attention map, which adaptively enhances salient features. 
While RAA blocks perform region-adaptive weighting on input feature maps $\boldsymbol{f}$, the feature representation capabilities at both global and local levels require further enhancement. To address this, we propose the FSCA block. This block strengthens feature representation for different regions by combining the FFT in the frequency domain with convolution representation in the spatial domain. Formally, the MGFE module can be expressed as follows: 
\begin{equation}
    \begin{aligned}
        \boldsymbol{f}^{\prime\prime} = \text{MGFE}(\boldsymbol{mask}, \boldsymbol{f}).
    \end{aligned}
\end{equation}

\textbf{Region-Adaptive Attention Block.} In the RAA block, we first employ a lightweight attention network to efficiently integrate the spatial priors of the mask into the encoded latent features $\boldsymbol{f}$. This network consists of two 3$\times$3 convolutional layers followed by a Sigmoid function. The objective of this process is to transform a hard binary mask (with values of 0 or 1) into a content-adaptive soft attention map $\boldsymbol{m}$. The convolutional operations learn to convert the precise pixel-level mask into region weights that are closer to the feature distribution and smoother, while the sigmoid function constrains these weights to the interval (0, 1). The generated attention map $\boldsymbol{m}$ is then multiplied element-wise with the original input feature map $\boldsymbol{f}$. This operation enhances features within the ROI region (where weights are close to 1) and suppresses features in the background region (where weights are close to 0). 
The RAA block can be formulated as
\begin{equation}
    \begin{aligned}
        & \boldsymbol{m} = \text{Sigmoid}(\text{Conv}_{3\times3}(\text{Conv}_{3\times3}(\boldsymbol{mask}))) ,\\
        & \boldsymbol{f}^{\prime} = \boldsymbol{f} \times \boldsymbol{m} + \boldsymbol{f}.
    \end{aligned}
\end{equation}

It is noteworthy that for the background decoder, the areas requiring enhancement are the inverse of those in the foreground decoder. The RAA block is modified as follows:
\begin{equation}
    \begin{aligned}
        \boldsymbol{f}^{\prime} = \boldsymbol{f} \times (\boldsymbol{1}-\boldsymbol{m}) + \boldsymbol{f}.
    \end{aligned}
\end{equation}

\textbf{Fequency-Spatial Collaborative Attention Block.} Following region-adaptive weighting by the RAA block, the FSCA block captures both global and local feature correlations from the feature map $\boldsymbol{f}^{\prime}$. It effectively leverages their complementary characteristics to enhance feature representation, as illustrated in Fig.~\ref{fig:MGFE}a. In contrast to conventional spatial-only attention, the FSCA block integrates both frequency-domain and spatial-domain attention to collaboratively enhance key features across two orthogonal dimensions, thereby facilitating high-quality image representation. The FSCA block first takes the output features $\boldsymbol{f}^{\prime}$ of the RAA block as input. The input is then processed in parallel through both the frequency attention (FA) branch $FreqAtt(\cdot)$ and the spatial attention branch (SA) $SpatAtt(\cdot)$. The outputs from these two branches are concatenated (Concat) along the channel dimension. 
A 1$\times$1 convolutional layer subsequently performs feature fusion and dimensionality reduction on the concatenated results. The fused features are then fed into a channel attention (CA) block to adaptively recalibrate the weights of different feature channels. 
Finally, attention features are added to the input $\boldsymbol{f}^{\prime}$ of the FSCA block through a residual connection to obtain the final feature maps $\boldsymbol{f}^{\prime\prime}$.
\begin{equation}
    \begin{aligned}
        &\boldsymbol{f}_{att} = \text{Concat} \left(\text{FreqAtt}(\boldsymbol{f}^{\prime}), \text{SpatAtt}(\boldsymbol{f}^{\prime}) \right), \\
        &\boldsymbol{f}^{\prime\prime} = \boldsymbol{f}^{\prime} + \text{CA} \left(\text{Conv}_{1\times1} \left( \boldsymbol{f}_{att} \right) \right).
    \end{aligned}
\end{equation}

For frequency attention blocks, as shown in Fig.~\ref{fig:MGFE}b, we adopt the approach proposed by~\cite{yu2022frequency,li2023embedding} by applying the FFT to decompose the input feature maps $\boldsymbol{f}^{\prime}$ into amplitude $f_{amp}$ and phase $f_{pha}$ spectrum. 
The amplitude spectrum represents the energy of different frequency components, while the phase spectrum contains structural and positional information~\cite{yu2022frequency}. We utilize the amplitude spectrum as guidance to adaptively modulate and optimize the phase spectrum. To achieve this, we primarily employ average pooling and a Softmax function to the amplitude spectrum to generate channel-wise frequency attention. The attention-weighted phase spectrum is then non-linearly transformed through two 1$\times$1 convolutional layers. Finally, the enhanced phase spectrum and amplitude spectrum are recombined and transformed back into the spatial domain using the inverse fast Fourier transform (IFFT).

For spatial attention blocks, as shown in Fig.~\ref{fig:MGFE}c, which operate in parallel to FA blocks, the focus is on capturing and enhancing local features in the spatial dimension. The core of the SA block is a multi-scale residual block (MRB) that captures multi-scale spatial context by fusing convolutional branches with differing receptive fields (e.g., 3$\times$3 and 11$\times$11). The input feature maps$\boldsymbol{f}^{\prime}$ are first processed through a series of stacked residual bottleneck blocks (RBs)~\cite{cheng2020learned} and embedded MRBs. A pixel-level spatial attention map is then generated using a Sigmoid function. Each value in this map ranges between 0 and 1, which adaptively enhances the spatial features of interest.

\subsection{Training Loss}
The objective of ROI-based image compression is to minimize the weighted sum of rate and distortion, where distortion is evaluated separately for foreground and background, with a higher penalty assigned to the foreground. To ensure high-fidelity reconstruction of the foreground while maintaining satisfactory visual quality in the background, we employ region-specific loss functions during training. The mean square error (MSE) loss for each region is defined as
\begin{equation}
    \begin{aligned}
        & \boldsymbol{D}_{FG} = \text{MSE}_{ROI}\left(\boldsymbol{x}\times \boldsymbol{mask} , \boldsymbol{\hat{x}}_{f} \times \boldsymbol{mask} \right), \\
        & \boldsymbol{D}_{BG} = \text{MSE} \left( \boldsymbol{x}, \boldsymbol{\hat{x}}_{b} \right). \\
    \end{aligned}
\end{equation}

To ensure the final fused images exhibit strong consistency and naturalness, we do not use a mask to separate the background when applying the MSE loss to the background decoder. This is because the reconstructed background region suffers from substantial information loss. Therefore, following~\cite{gfpgan}, we incorporate both style loss and perceptual loss to enhance its visual quality.
\begin{equation}
\begin{aligned}
    & \boldsymbol{L}_{per} =  \mathbb{E} \left( \sum^N_{i=1}  \left\Vert \psi_{i} \left(  \boldsymbol{x} \right) - \psi_{i} \left( \boldsymbol{\hat{x}}_{b} \right) \right\rVert_2 \right) ,\\
    & \boldsymbol{L}_{sty} =  \mathbb{E} \left( \sum^N_{i=1}  \left\Vert \text{Gram}\left(\psi_{i} \left(  \boldsymbol{x} \right)\right) - \text{Gram}\left(\psi_{i} \left( \boldsymbol{\hat{x}}_{b} \right)\right) \right\rVert_2 \right) ,\\
    & \boldsymbol{L}_{b{\_}{rec}} = k_1 \cdot \boldsymbol{L}_{per} + k_2 \cdot \boldsymbol{L}_{style}, 
\end{aligned}
\end{equation}
where $\psi_{i}(\cdot)$ denotes the $i$-th convolutional layer of the pre-trained VGG network~\cite{simonyan2014very}. $N$ is the total number of intermediate layers used for feature extraction. $\text{Gram}(\cdot)$ denotes the Gram-matrix function employed in the style-feature loss. 

ROI-based image compression is still essentially an RDO problem. By referring to Eq.~\ref{eq:optimization}, the whole loss function proposed in this article is defined as
\begin{equation}
    \begin{aligned}
        & \boldsymbol{\mathcal{R}} = \mathbb{E} \left[ -\log_2 p_{\hat{y}} (\boldsymbol{\hat{y}}|\boldsymbol{\hat{z}}) \right] + \mathbb{E} \left[ -\log_2 p_{\hat{z}} (\boldsymbol{\hat{z}}) \right],\\
        & \boldsymbol{\mathcal{L}} = \lambda \cdot \boldsymbol{\mathcal{R}} + (\boldsymbol{D}_{FG} + k_3 \cdot \boldsymbol{D}_{BG} ) + \boldsymbol{L}_{b{\_}{rec}},  \\
    \end{aligned}
\end{equation}
where $\lambda$ is a trade-off parameter to balance rate and distortion.  $p_{\boldsymbol{\hat{y}}}$ and $p_{\boldsymbol{\hat{z}}}$ are the corresponding probability distribution of $\boldsymbol{\hat{y}}$ and $\boldsymbol{\hat{z}}$. The hyperparameters $k_1$, $k_2$, and $k_3$ are set to 0.1, 0.02, and 50, respectively.
\begin{figure*}[!t] 
	\centering    
	\includegraphics[scale=0.243]{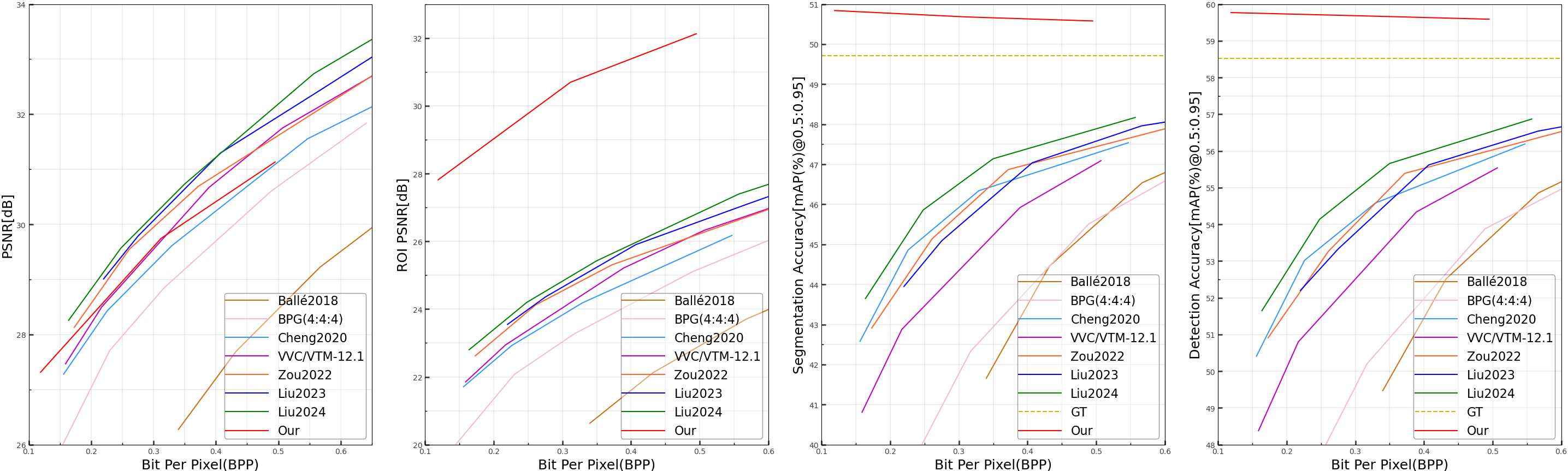} 
	\caption{RD performance averaged in terms of PSNR, ROI-PSNR, and mAP@0.50:0.95 on COCO2017 dataset.}
	\label{fig:psnrsegbpp}
\end{figure*}

\section{EXPERIMENTS}
\subsection{Experimental Settings}
Our model is trained on the training subset and evaluated on the validation subset of the COCO2017 dataset~\cite{lin2014microsoft}. We randomly select 15,000 images and crop them to a size of 256$\times$256 for training. The masks used in our experiments are generated by applying pixel-level binarization to the semantic-segmentation annotations provided with the dataset. We specifically select image-mask pairs where the mask-to-image pixel ratio falls within the range of $[0.08,0.8]$ for both training and testing. Our model is optimized with Adam optimizer for 450 epochs with a batch size of 16. We employ a learning rate decay strategy: the learning rate is initialized at $1\times10^{-4}$ and reduced to $5\times10^{-5}$ for the last 100 epochs. The Lagrange multipliers $\lambda$ are set to $\{64, 128, 512\}$. The channel parameters N and M is set to 192 and 320, respectively. All experiments are conducted on a server equipped with an NVIDIA GeForce RTX 4090D GPU. 

\begin{figure*}[!t] 
	\centering    
	\includegraphics[scale=0.62]{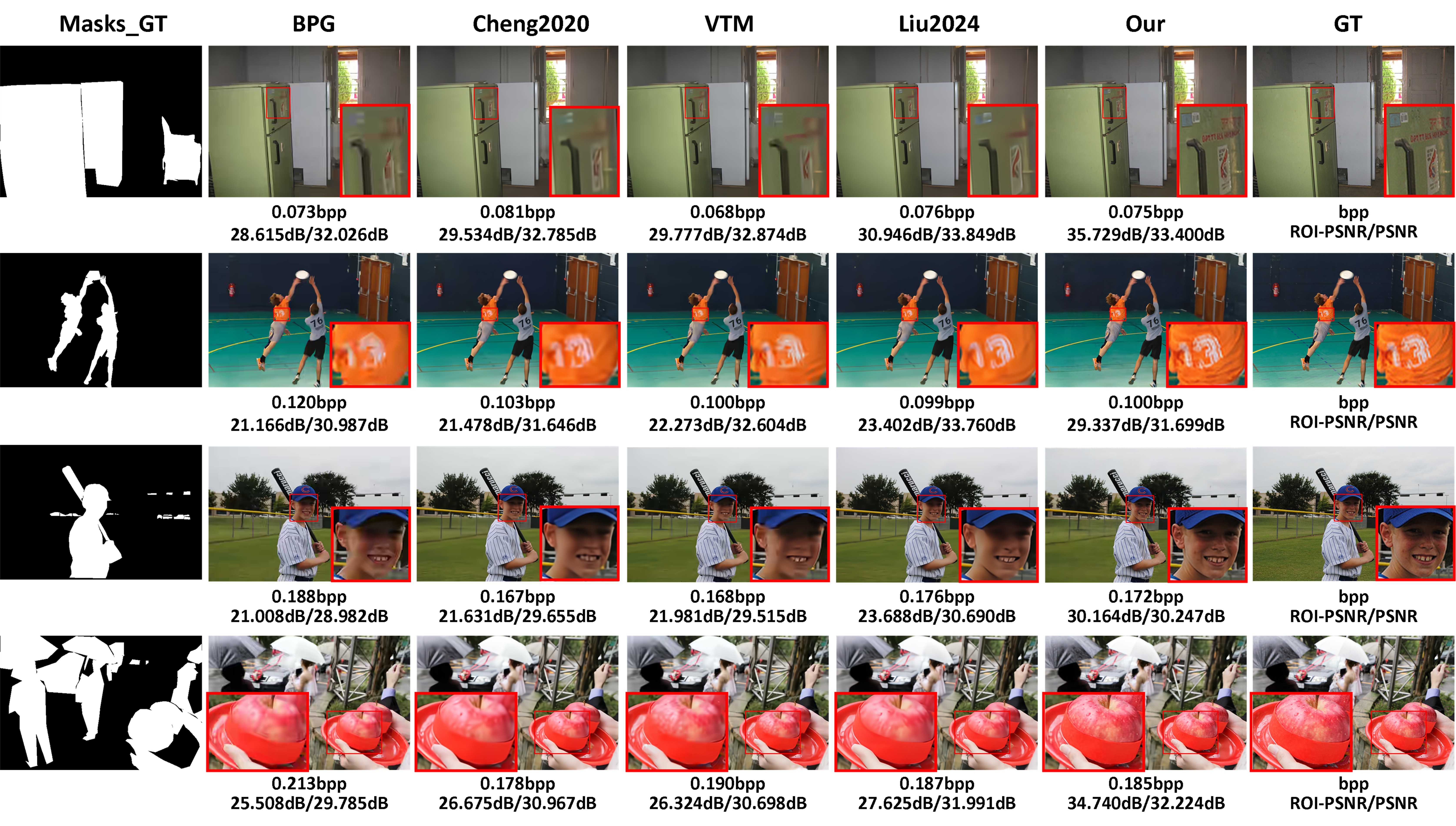} 
	\caption{Comparisons on decoded images with different compression methods. Zoom in for best view.}
	\label{fig:psnrbppvisual}
\end{figure*}
\subsection{Performance Analysis of Image Compression} 
To evaluate the coding performance of our proposed method, we compare it with several state-of-the-art LIC methods, including those by Ballé2018~\cite{balle2018variational}, Cheng2020~\cite{cheng2020learned}, Zou2022~\cite{zou2022devil}, Liu2023 (TCM-LIC)~\cite{liu2023learned}, and Liu2024~\cite{liu2024region}. We adopt the only small model in Liu2023 (TCM-LIC)~\cite{liu2023learned} with fully open pre-trained weights for experimental comparison. Furthermore, we include comparisons with classic hybrid codecs, namely BPG(YUV444)~\cite{bpg} and VVC/VTM-12.1 Intra(YUV444)~\cite{vtm}. The compression rate is measured in bits per pixel (BPP), and the reconstruction quality is evaluated using the peak signal-to-noise ratio (PSNR).
\begin{table} [!ht]
  \centering
 \renewcommand{\arraystretch}{2.0}
  \caption{Average BD-rate ($\%$) gain and BD-mAP ($\%$) gain compared to VVC across COCO2017 dataset. Symbol (``-'') indicates better performance than VVC for BD-rate gain, while indicates the opposite for BD-mAP gain.}
  \label{tab:mapBD}
  \scalebox{0.66}{
    \small
      \begin{tabular}{lccccc}  \hline
         Methods &  ROI-PSNR & \multicolumn{2}{c}{Segmentation Accuracy} & \multicolumn{2}{c}{Detection Accuracy} \\ 
           & BD-rate(\%)  & BD-rate(\%) & BD-mAP(\%)  & BD-rate(\%) & BD-mAP(\%) \\ \hline
         VVC  & --  & -- & -- & -- & --  \\ 
         Ballé2018~\cite{balle2018variational} &  116.75  & 57.10 & -3.034 & 52.01 & -2.638 \\ 
         BPG(YUV444)  & 31.39  & 47.97 & -3.676 & 45.79 & -3.475 \\ 
         Cheng2020~\cite{cheng2020learned} &  8.02  &  -22.93 & 1.591 & -24.32 & 1.515 \\ 
         Zou2022~\cite{zou2022devil}  & -9.54  & -19.60 & 1.374 & -20.30 & 1.471 \\
         Liu2023~\cite{liu2023learned}  & -10.05  & -14.97 & 0.973 & -17.95 & 1.108 \\
         Liu2024~\cite{liu2024region}&  -17.12  &  -30.09 & 2.135 & -32.04 & 2.119  \\ 
         Explicit bit allocation &  -66.80  &  -70.01 & 5.338 & -55.42 & 5.303  \\ 
        \textbf{Our} & \textbf{-84.67} & \textbf{-99.88} & \textbf{7.472} & \textbf{-99.98} & \textbf{7.279} \\ \hline
    \end{tabular}}
\end{table}
Specifically, we obtain the experimental results for the competing methods primarily by conducting tests on the CompressAI platform~\cite{begaint2020compressai} or the pre-trained model provided by the authors. The overall compression performance is evaluated using the BD-rate metric, which is computed with the Bjøntegaard algorithm~\cite{Bjontegaard}. 
A negative BD-rate value indicates superior compression performance, meaning that a lower bitrate (bpp) is required to achieve the same PSNR compared to the anchor. Conversely, a positive BD-rate value indicates inferior performance.

To have a more intuitive understanding of the overall coding performance among different compression methods, we plot their RD curves for PSNR, ROI-PSNR, and BPP. The experimental results are shown in Fig.~\ref{fig:psnrsegbpp}. The results demonstrate that our method achieves significantly superior coding performance for the image's ROI. Furthermore, while the global coding performance of our method is slightly lower than the latest Liu2024 method~\cite{liu2024region}, it remains considerably better than the classic Cheng2020 method~\cite{cheng2020learned}. To quantitatively assess ROI coding performance, we employ the BD-rate metric. For a fair comparison, we use the classic VVC standard~\cite{vtm} as a baseline. The corresponding results are summarized in Table~\ref{tab:mapBD}. The data show that LIC methods generally outperform traditional codecs. Notably, the Liu2024 method~\cite{liu2024region} achieves a BD-rate reduction of 17.12\% compared to VVC. In contrast, our proposed ROI-based method achieves a substantially larger BD-rate reduction of 84.67\%, attaining the best overall RD performance.

In addition to quantitative comparisons, we provide a qualitative analysis of different compression methods. The results are presented in Fig.~\ref{fig:psnrbppvisual}, which visualizes decoded images from various methods at similar bitrates to facilitate an intuitive comparison of reconstruction quality in both ROI and non-ROI regions.
Our method achieves minimal information loss in ROI areas, with reconstructed images closely approximating the original uncompressed images (GT). It preserves relatively rich texture details even at lower bitrates.

Although the Liu2024 method~\cite{liu2024region} achieves superior RD performance globally, it suffers from noticeable information loss in ROI regions. Furthermore, compared to the Cheng2020 approach~\cite{cheng2020learned}, our method not only delivers high fidelity in ROIs but also maintains better visual quality in background regions. These advantages are primarily attributed to our implicit bit-allocation method and the proposed dual-decoder architecture. In conclusion, our ROI-based compression method achieves optimal RD performance while simultaneously maintaining satisfactory visual quality in background areas.

\begin{figure*}[!t] 
	\centering    
	\includegraphics[scale=0.36]{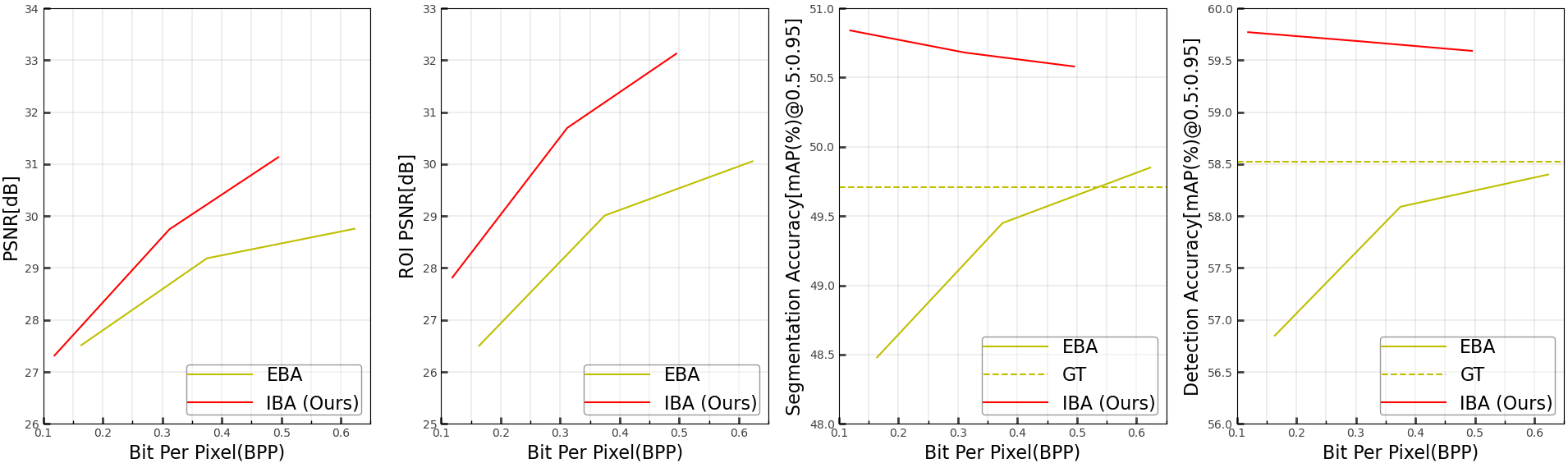} 
	\caption{Performance comparison of different bit allocation methods. ``EBA'' and ``IBA'' represent explicit and implicit bit allocation methods, respectively. }
	\label{fig:bitpsnrbpp}
\end{figure*}
\subsection{Performance Analysis of Machine-vision Tasks}
To evaluate the impact of different compression methods on the performance of downstream computer-vision (CV) tasks, we employ the YOLOv11 model~\cite{khanam2024yolov11} to perform object detection and instance segmentation tasks on compressed images. The YOLOv11 provides multiple pre-trained models suitable for different tasks. We select the mean average precision with an IoU threshold of 0.5:0.95 (mAP@0.50:0.95) as the primary evaluation metric. Using this metric, we quantitatively assess the segmentation and recognition accuracy of different compression methods via the BD-rate and BD-mAP metrics.
A negative BD-rate value indicates that fewer BPP are required to achieve the same mAP, representing superior compression performance. Conversely, a positive BD-rate value indicates inferior performance. Similarly, a positive BD-mAP value signifies a higher mAP at the same bpp, denoting better performance, while a negative value indicates the opposite.

The experimental results in Fig.~\ref{fig:psnrsegbpp} demonstrate that our compressed images achieve optimal recognition accuracy for both instance segmentation and object detection tasks, significantly outperforming the GT-test results. While Liu2024~\cite{liu2024region} method yields relatively good recognition accuracy, its performance remains substantially inferior to the GT. Notably, our method's superior accuracy compared to the GT primarily stems from its high-fidelity preservation of foreground elements combined with the suppression of high-frequency textures in background regions, which effectively reduces interference from non-essential information. Furthermore, the slight decrease in recognition accuracy observed with increasing BPP can be attributed to added texture details in background regions at higher bitrates, which may introduce additional complexity to the recognition tasks.

As shown in Tab.~\ref{tab:mapBD}, our method achieves a 7.472\% improvement in BD-mAP@0.50:0.95 alongside a 99.88\% reduction in BD-rate for the instance segmentation task. Similarly, for object detection, it demonstrates a 7.279\% gain in BD-mAP@0.50:0.95 and achieves a 99.98\% BD-rate reduction. These results conclusively demonstrate that our method not only delivers substantial performance gains but also effectively enhances recognition accuracy in CV tasks.

 \begin{figure*}[!t] 
	\centering    
	\includegraphics[scale=0.525]{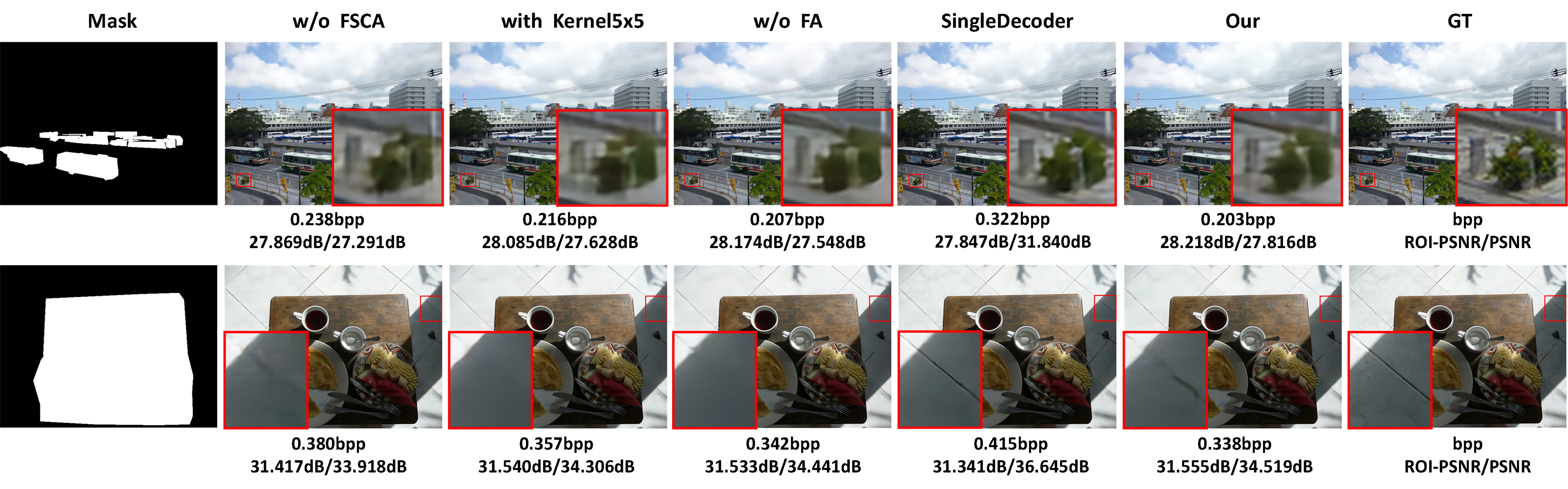} 
	\caption{Comparison of visualization results for different variants in our model. Zoom in for best view. }
	\label{fig:ablatvisual}
\end{figure*}
\subsection{Performance analysis of different bit allocation}
To ensure a fair experimental comparison, the model for the explicit bit-allocation method is constructed using the backbone of our proposed model (excluding the RAA block and the background decoder), while adopting the explicit bit allocation and training strategy from~\cite{cai2019end}. First, to compare and analyze the RD performance of the two bit-allocation methods, we plot their RD curves for PSNR, ROI-PSNR, and BPP, as shown in Fig.~\ref{fig:bitpsnrbpp}. The results demonstrate that our method outperforms the explicit approach in both ROI and global image quality, with the performance gap being particularly pronounced in ROI coding. We further quantify the ROI coding performance using the BD-rate metric (Table~\ref{tab:mapBD}). Although the explicit method achieves a 66.80\% BD-rate reduction compared to VVC, our method yields a significantly greater improvement of 84.67\%.
Additionally, we visualize the latent feature maps and bit-allocation maps (Fig.~\ref{fig:Im_explivisual}). The visualization reveals that the hard-gating mechanism in the explicit bit-allocation method disrupts the statistical distribution of the entropy model and impairs the predictive capability of the spatial auto-regressive context model. This leads to suboptimal bit allocation in ROI regions, with an unnecessarily high-bit consumption.

To comprehensively evaluate the impact of different bit-allocation methods, we also employ CV tasks for analysis. As demonstrated in Fig.~\ref{fig:bitpsnrbpp}, our method achieves superior performance in both instance segmentation and object detection tasks, significantly outperforming both the explicit approach and the GT-test results.
We subsequently quantify segmentation and recognition performance using BD-rate and BD-mAP metrics. The data in Table~\ref{tab:mapBD} show that while the explicit method achieves BD-rate reductions of 70.01\% and 55.42\% for instance segmentation and object detection, respectively, along with BD-mAP gains of 5.338\% and 5.303\%, these results remain substantially inferior to ours. This further validates the superior performance of our implicit bit-allocation approach and its significant advantage in facilitating efficient CV task implementation.

\subsection{Ablation Studies}
To verify whether our method achieves optimal overall performance, we conduct an ablation study on key components of the backbone model. The evaluated variants are: (1) removal of the FSCA block from the MGFE module (w/o FSCA), (2) removal of the FA component from the FSCA block (w/o FA),  (3) removal of the background decoder and use only a foreground decoder (SingleDecoder), and (4) replacement of the 11$\times$11 convolutional kernel in the SA block with a smaller 5$\times$5 kernel (with kernel5$\times$5).
\begin{figure}[!t] 
	\centering    
	\includegraphics[scale=0.15]{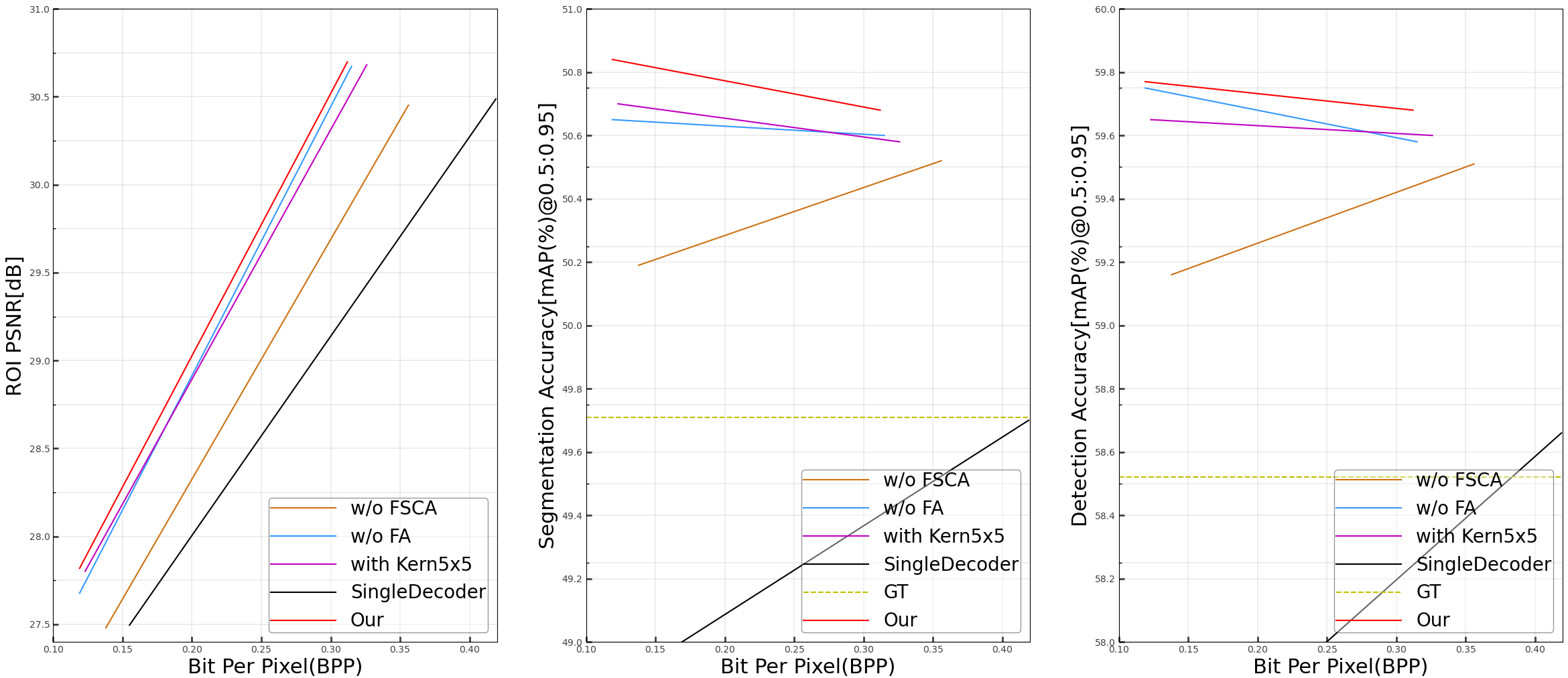} 
	\caption{Performance comparison of different variants in terms of ROI-PSNR and mAP@0.50:0.95 evaluation metrics.}
	\label{fig:abltion}
\end{figure}
To assess the impact of these variants, we plot their RD curves for ROI-PSNR and BD-mAP against BPP, as shown in Fig.~\ref{fig:abltion}. The results indicate that the use of the single decoder and the removal of the entire FSCA module considerably degrade the coding performance for the image ROIs and significantly reduces the recognition accuracy in the downstream CV tasks. Furthermore, the variant with the 5$\times$5 kernel (compared to 11$\times$11) demonstrates that a substantially smaller receptive field weakens the model's ability to capture local high-frequency information, particularly in ROI regions. This impairment not only compromises overall coding performance, but also reduces CV task accuracy. In contrast, the FA component, which captures global feature dependencies in the frequency domain, is shown to be a critical factor in enhancing our method's performance.

We further visualize the compression results of the different variants to examine their impacts in detail, as depicted in Fig.~\ref{fig:ablatvisual}. Visual comparisons confirm that omitting the FSCA module substantially degrades performance, both by reducing reconstructed image quality and increasing bit consumption. Employing the smaller kernel weakens the model's ability to represent fine textural details, whilst also increasing BPP occupancy. When employing only a single decoder, the coding network fails to achieve a balance between high fidelity in ROI and visual quality in background regions. Although the global image exhibits satisfactory perceptual quality, this variant results in higher BPP consumption and limited ROI fidelity. In summary, the FSCA module and its components are crucial for the coding performance of our model. They are instrumental in enhancing the reconstruction of textual details and minimizing bitrate consumption.

\section{CONCLUSION}
In this paper, we propose an efficient ROI-based deep image compression method that employs an implicit bit-allocation strategy.This approach addresses the limited compression performance of existing explicit methods, which arises from their disruption of the entropy model's statistical distribution. To apply ROI masks for implicit bit allocation, we introduce an RAA block to generate smooth and differentiable attention maps from binary input masks. We further propose a FSCA block to enhance both global and local representations of the weighted feature maps. Building upon the introduced RAA and FSCA blocks, we develop a novel MGFE module to progressively guide layer-wise feature learning in the coding network. To simultaneously achieve high ROI fidelity and well-preserved background quality, we employ a dual-decoder architecture that separately reconstructs the foreground and background regions. Experimental results demonstrate that our method achieves state-of-the-art RD performance while maintaining an optimal balance between high-fidelity ROI reconstruction and background preservation. Furthermore, we provide a detailed comparative analysis of explicit and implicit bit-allocation methods. The experimental evidence confirms that our implicit-based approach not only delivers superior coding performance but also significantly improves the accuracy of downstream CV tasks. Notably, our compressed images even yield higher recognition accuracy than the original uncompressed images.

\bibliographystyle{IEEEtran}
\bibliography{reference}

\newpage




\vfill

\end{document}